# Merit or networks? What decides where research is published

Ning Li (Tsinghua University)
lining@sem.tsinghua.edu.cn
May 2026.

## Significance Statement

Modern science runs on prestige: where work is published shapes who gets read, funded, and hired. A persistent worry is that these decisions reward who you know, not the quality of the work. It has been hard to settle because a paper's quality could not be judged ahead of its fate without using that fate as the yardstick. By scoring a paper's idea from its text, before publication and independently of its authors, we separate what the work itself signals from the advantage of connections. A clear structure emerges: execution mainly governs whether a paper is published, idea quality grades the ladder, and connections matter most at the top. Publishing is neither pure meritocracy nor pure network, but a measurable mix.

## Abstract

Does scientific publishing reward the quality of ideas or the advantage of connections? The question is universal to prestige-driven science, yet it has resisted decades of study because a paper's quality could not be gauged ahead of its publication fate without using that fate as the yardstick. We break this constraint by measuring a paper's idea quality directly from its text, before publication, using a discipline-trained evaluator that scores the idea without seeing author names or outcomes. Using economics as a case study, we combine this text-legible idea-quality score with an execution-quality rubric, a connection index, an author-ability index, and an off-the-shelf language-model text score to estimate a five-input production function for journal placement across 6,208 economics working papers. The inputs form a sequence along the ladder of prestige. Execution sets a meritocratic floor and is the largest input overall. Text-legible idea quality grades the rungs in between. Connections set a favoritism ceiling that bites mainly near the apex, the most selective journals. Connections work through two additive channels: connected authors write papers that score higher, and at equal scores their papers are still more likely to place better. Yet this advantage is bounded. Connections raise the odds of every rung without making the apex the typical outcome for ordinary ideas, and even the highest-scoring papers face real friction reaching the visible journal ladder. The result nests, rather than chooses between, the meritocracy and network accounts of how science is published.

## Introduction

Every scientific field runs a prestige tournament (1). A small number of journals, awards, grants, and faculty posts confer outsized visibility on the work they accept, and where a finding lands shapes what gets read, funded, and built upon. Two accounts of how those decisions are made have long competed. In the first, the tournament is a meritocracy: strong work, rigorously executed, finds its level, and the most selective venues are selective because they choose the best. In the second, the tournament is a network: gatekeepers favor the people they already know, prestige compounds on prestige (2), and where a paper lands reveals as much about its authors' connections as about its content. The stakes are broad. If gatekeeping rewards networks over substance, science misallocates its scarcest resource (attention), and the misallocation is invisible precisely where it matters most, at the top.

The debate has resisted resolution for a reason that is methodological, not ideological. To ask whether connections matter beyond the quality of the work, one must hold quality fixed. But the only measures of quality available before a paper's fate is decided (the authors' reputation, their prior citation counts, the prestige of their institution) are themselves saturated with the very network one is trying to net out. And the cleanest measure available after the fact, where the paper was ultimately published, is the outcome. Researchers have therefore had to choose between a contaminated pre-treatment control and a circular post-treatment one, leaving a literature rich in evidence yet inconclusive on the central question. Studies of gatekeeper ties show that connected authors publish more, and in better venues, and that connections can shape the treatment a paper receives in review (3–7). Several of these works do try to separate favoritism from quality, but only with ex-post citations, which are themselves shaped by author status, publication venue, and a documented bias against novel work (8–10), and so cannot hold pre-publication quality fixed. A separate strand distinguishes the importance of a paper's idea from the rigor of its execution, and shows that the quality of the writing is not incidental but predicts how a paper fares in review (11, 12). Experiments that randomize author identity show that single-blind review confers an advantage on famous and elite-institution authors. This is direct evidence that names, not only content, move gatekeeping decisions (13). What no prior measure could supply is a quality signal formed before the outcome and free of the author's name; that is the gap we fill.

We break the impasse with a recently developed instrument: a large language model fine-tuned on a discipline's own publication record, so that it scores a paper's idea by the standard a field's accumulated judgment has set (14). It belongs to a fast-growing class of language-model evaluators of scientific text (15–17), but unlike off-the-shelf scorers it is tuned to a discipline's own realized judgments, and that is what makes it accurate where general-purpose models are not. On held-out benchmarks it recovers the field's four-tier judgment with roughly 70% accuracy in economics, far above the 25% chance baseline and the ~30% reached by frontier reasoning models. Because the model reads only text, with no author names, institutions, or venue, its score reflects what the work says rather than who wrote it. That gives the favoritism literature the pre-publication, name-blind measure of quality it has always lacked, one not built out of the network it is meant to net out. Our subject is the publication game itself, and the measure simply lets us hold quality fixed and ask what connections add conditional on that text-legible idea quality and a separate execution rubric.

We study this general phenomenon in one well-instrumented case: economics. The question needs two ingredients, and economics supplies both. Its pre-publication text is public, circulated as working papers, so a name-blind ex-ante quality signal can actually be built. And its prestige ladder is a matter of clear disciplinary consensus: the "Top-5" general-interest journals and the leading top-field journals are widely agreed upon (18), so the outcome is measurable and the apex is unambiguous (Methods makes this concrete). For 6,208 economics working papers circulated as preprints between 2010 and 2015, we assemble five measured inputs for each paper: the text-legible idea-quality signal just described; an independent rubric score for execution (the quality of the writing, identification, methods, robustness, mechanism, and external validity); a measure of the authors' professional connections, built strictly from information dated before the paper appeared; a measure of author ability (prior productivity and track record); and, as a contamination-free anchor, the rating a general-purpose language model gives the same text. We then estimate a five-input "production function" for where each paper is ultimately published, mapping outcomes onto a five-rung prestige ladder that runs from unpublished up to the field's five most prestigious journals, its apex venues, conventionally called the "Top-5." With all five inputs observed at once on the same pre-publication population, the two stories stop being rival worldviews and become competing, testable restrictions on one model. A pure meritocracy predicts that the residual connection gradient vanishes once quality is observed, that connections do not interact with quality, and that the highest-rated papers are essentially never left unpublished. A pure network predicts a large connection gradient at every rung, a strong multiplicative interaction (connections "unlock" good ideas), and an unbounded "elevator" that carries weak work to the apex. Neither set of predictions survives contact with the data, and the pattern of which predictions fail is the contribution.

Four findings follow, and they cohere into one picture. The first is that the inputs form a sequence along the ladder rather than competing for one spot. In an order-invariant decomposition of the explained variance in prestige placement, execution is the single largest input, accounting for 37.3 percent, more than connections (24.2 percent), the text idea-quality signal (16.4 percent), author ability (13.9 percent), or the general-purpose model (8.2 percent). This ranking is not constant across the ladder. The share attributable to connections rises monotonically with selectivity, from 15.3 percent of explained variance at the "published at all?" margin to 38.8 percent at the apex (Top-5) margin, while the execution share moves inversely, dominating the broad margins and ceding ground to connections at the very top. Gatekeeping by connection is real, and it is concentrated exactly where the stakes are highest, but it is not the largest force in the system as a whole.

The second finding is the center of gravity of the paper: connections operate through two distinct channels that add rather than multiply, and the favoritism channel is bounded. The first channel is capture: connected authors genuinely write papers that score higher on the text signals (the mean idea-quality score rises by about 0.6 standard deviations from unconnected to highly connected authors). The second is favoritism: holding the text-legible idea signal fixed, connected papers still place better. Among papers the evaluator rates only "fair," the apex (Top-5) rate climbs from 1.9 percent for the unconnected to 14.1 percent for the highly connected, a gap that equal text-legible quality cannot explain. These two channels are additive, not interactive: the quality-by-connection interaction is economically negligible and statistically insignificant (interaction coefficient −0.007, $p = 0.20$). Connections do not "unlock" good ideas; they shift placement up by a roughly constant amount on top of whatever the idea signal delivers. And the favoritism they buy is bounded. Connections raise the odds of every rung, but where the added mass lands depends on the idea tier. A fair-rated, highly connected paper can reach the apex, and 14.1 percent do, yet its connections lift it mostly into the next rung down (top-field placement rises to 30.3 percent). The apex does not become its typical destination; the balance tips into the apex only for ideas the signal already rates at the top. In this sense the folk claim "it's who you know" and the indignant reply "no, it's the work" are both correct and describe different parts of one machine.

The third finding is that even the work the signal ranks highest faces real friction reaching the visible ladder. If publishing were frictionless for the papers a discipline-trained signal loves, the very top of the score distribution would almost never go unpublished. But it does. Among the top one percent of papers by the idea-quality score, 57.1 percent reach the apex, yet 17.5 percent are never matched to any tier of the journal ladder at all (against 34.9 percent in the bottom half). The signal sharply predicts the ceiling, a roughly tenfold spread in apex rates across the distribution, but only weakly protects against the floor, a roughly twofold spread in the rate of going unpublished. High text-legible quality buys a shot at the top; it does not buy insurance against falling through the cracks (19).

The fourth finding is that these facts assemble into a single structure: a meritocratic floor, a favoritism ceiling, and an idea signal that grades the ladder in between. Execution sets the meritocratic floor: among papers with low text-legible idea quality, strong execution cuts the unpublished rate from 46.9 to 20.6 percent and raises the rate of reaching top-field-or-better from 11.8 to 48.1 percent, while weak execution leaves even high-idea-quality papers unpublished 40.1 percent of the time. Connections set the favoritism ceiling: they add almost nothing at the publish-or-not margin (the unpublished rate is nearly flat across connection tiers) but a large premium at the apex. The text-legible idea signal grades the rungs between. The publication game is neither a pure meritocracy nor a pure network; it is a meritocratic floor with a favoritism ceiling, and most of the work in the middle is done by execution.

This structure carries three contributions. Substantively, it provides the first joint accounting of ideas, execution, connections, and ability for a large, legally clean preprint cohort, and shows that decades of "favoritism regressions" were misspecified in a specific, correctable way: because execution is the largest input and is correlated with connections, any connection coefficient estimated without an execution control silently absorbs unmeasured execution quality. Methodologically, it demonstrates that an ex-ante, text-derived, name-blind quality measure can serve as the missing pre-publication control that this literature has always lacked, and it shows how to use such a measure responsibly given that it is trained on the outcome (replace it with an untrained model; compare within authors; bound the rest). Conceptually, it offers a portable vocabulary, including factor shares, marginal returns, a bounded elevator, and apex friction, for any prestige market with text-legible inputs and a measurable outcome ladder, from grant funding to hiring to conference admission.

The remainder of the paper follows standard order. Methods details the data sources and sample construction, the five measured inputs, the estimating equation and the quantities estimated, and the identification strategy, stated as a single assumption with every robustness check organized as a bound on a named threat to it. Results then reports the factor-share sequence, the dual (capture-plus-favoritism) channel and its boundedness, the contrasting roles of execution and connections at the floor and the ceiling, the divergence between what wins placement and what wins citations, and apex friction. Discussion reconciles the meritocracy and network views as two channels of one machine, draws out the methodological lesson that execution is the omitted variable in prior work, weighs whether the bounded favoritism gradient reflects relationship rents or soft information, and states the welfare and portability limits of the design. Supplementary Information collects the most technical detail: the latent placement-index model and its meritocracy-versus-network restrictions, the sample-construction steps, the multinomial and per-dimension decompositions, the full threat-by-threat sensitivity stack, the alternative-outcome and specification-curve checks, and the heterogeneity of the dual channel across fields, teams, and time.

## Methods

### Study system and sample

Economics is an unusually clean test bed for studying scientific gatekeeping, for two concrete reasons. First, the pre-publication text is publicly available: economists circulate their work as working papers before it is judged by journals, so an ex-ante, name-blind quality signal can be built from the same text a paper presents to the market. This is the enabling measurement the question has always lacked. Second, the prestige ladder has clear disciplinary consensus. The field agrees on a small set of apex general-interest journals (the "Top-5": *American Economic Review*, *Quarterly Journal of Economics*, *Journal of Political Economy*, *Econometrica*, and *Review of Economic Studies*) and on a recognizable set of leading specialty ("top-field") journals, so the outcome, where a paper lands, is measurable and the apex is unambiguous. Most fields lack one or both: either no public pre-publication text from which a quality signal could be read, or no widely agreed-upon prestige hierarchy against which placement could be ranked. Economics has both, which is what makes the production-function measurement here possible at all. We therefore treat the economics publishing market as a tractable instance of a general problem: a prestige hierarchy of journals in which a paper's destination is determined by some combination of its content and its authors' position in the social network of gatekeepers.

The source population is working papers circulated by the National Bureau of Economic Research (NBER), a leading preprint series in economics. We use NBER rather than other repositories for both legal and practical reasons: it provides structured public metadata and stable identifiers, whereas some repositories' terms of use prohibit the text and data mining this project requires. We never scrape or mine any source whose terms forbid it.

The joined dataset contains 6,210 NBER working papers issued 2010–2015; 6,208 carry a usable text-legible idea-quality score. Two analytical samples appear throughout. Measurement and calibration exhibits use all 6,208 scored papers, because calibrating the idea-quality signal against realized placement does not require the connection and ability data. The five-input production-function estimates use a 5,848-paper complete-case sample (papers non-missing on all five inputs), comprising 3,067 primary-author clusters. Every reported number is tagged to one of these two bases, and the choice does not move the conclusions: calibration and percentile diagnostics computed on the 5,848-paper sample differ from the 6,208-paper versions only in the third significant figure. The unit of analysis is the paper, and standard errors are clustered on the primary author throughout. SI Table A0 reports the step-by-step sample construction. De-identified data and code that reproduce every table and figure are available at https://github.com/ln9527/publishing-production-function (SI, Data availability).

### The placement ladder

We matched publication outcomes from NBER metadata, version links in the OpenAlex bibliographic database, the Crossref registry, and conservative fuzzy title–author matching. We then mapped each paper to a five-rung placement ladder: 0 = unmatched to the four-tier journal panel; 1 = lower-tier; 2 = mid-tier; 3 = top-field (the leading specialty journals); 4 = Top-5 (the five elite general-interest journals named above). The headline binary outcome, prestige placement, equals one for top-field or Top-5. We also report Top-5 alone, any publication, the 0–4 linear ladder, and log citations as alternative outcomes. Privileging the prestige binary is a substantive choice: a multinomial decomposition (SI) shows the idea-quality signal is monotone and strongest at the top of the ladder but does not cleanly order "lower-tier versus unpublished," so the prestige binary is the most interpretable gatekeeping margin, and the conclusions hold across the alternative outcomes.

### The five measured inputs

We decompose what a paper brings to the publishing market into five inputs, each measured before or independently of the journal outcome.

Text-legible idea quality (q) is the predicted quality grade a discipline-trained evaluator (14) assigns to a distilled "pitch" of each paper, a structured summary of its research question and contribution. The evaluator reads text only and, at inference, sees no author names, institutions, or eventual placement. We use two independently trained checkpoints; the production-function regressor is their ensemble expected tier value, and the calibration exhibit uses a log-probability geometric-mean ensemble. The signal is well-calibrated at the top (SI Table S1): among papers that realized Top-5 placement, 53.3% carry an ensemble argmax grade of "exceptional" and 94.1% of "strong or exceptional," with a mean score of 3.36 for Top-5 papers versus 2.85 for unmatched papers.

Execution quality (X) is a full-paper rubric, produced independently of the idea score, scoring six dimensions: writing and presentation, identification, econometric methodology, mechanism and external validity, contribution, and robustness. The composite is close to unidimensional; per-dimension detail is reported in SI.

Connections (C) are built under strict temporal censoring: every connection, ability, and prestige feature uses only data dated before the paper's NBER issue date, because leakage of post-issue information would break identification. The block includes acknowledged editor links, prior editorship, coauthor proximity to editors, coauthor-network centrality (20), institutional prestige, mobility, and advisor links where available. The scalar measure is an inverse-covariance-weighted index (21); because connections are genuinely multidimensional (the first principal component explains only 31%), multi-indicator specifications appear as robustness.

Author ability (A) is separated from connections by construction: prior publications, prior Top-5 publications, recent productivity, academic age, and related pre-issue measures, collapsed to an index of the same form (21). The separation matters because much of what is colloquially called "connections" is in fact ability or career stage.

Off-the-shelf text score (L) is a general-purpose large language model's rating of the same pitch text (22), not trained on journal outcomes. It is the anchor free of outcome-training circularity: any pattern that holds when q is replaced by L cannot be an artifact of having trained the score on the journal tier. Off-the-shelf models do carry pretrained priors and, given identifying cues, can rate prestigious-looking work more highly (17); scoring a short, name-blind pitch limits those priors here, and the within-author check bounds what remains.

Two features of the input correlation matrix (Table 1) matter for what follows. The text signals and execution co-move (corr(q, X) = 0.36; corr(q, L) = 0.44): good ideas and good execution travel together, which is why omitting execution biases the idea coefficient. The connection index is only weakly correlated with the quality signals (corr(q, C) = 0.06; corr(X, C) = 0.06), so the favoritism residual is not mechanically small and the two connection channels are largely orthogonal.

Table 1. Summary statistics and correlations among the five measured inputs (analytical sample, N = 5,848).

*Panel A. Summary statistics (five-block analytical sample)*

| Variable | N | Mean | SD | Min | Median | Max |
|---|---|---|---|---|---|---|
| SFT predicted idea quality | 5848 | 3.039 | 0.558 | 1.005 | 3.145 | 3.979 |
| Off-shelf LLM text score | 5848 | 2.346 | 0.318 | 1.25 | 2.2 | 3.81 |
| Execution quality | 5848 | 3.711 | 0.81 | 1.15 | 3.85 | 5.0 |
| Connection index (Anderson) | 5848 | 0.0 | 0.325 | -1.186 | -0.003 | 1.663 |
| Author ability | 5848 | -0.0 | 1.384 | -4.863 | -0.369 | 15.037 |

| Variable | N | Mean | SD | Min | Median | Max |
|---|---|---|---|---|---|---|
| index (Anderson) | | | | | | |
| Team size | 5848 | 2.501 | 1.031 | 1.0 | 2.0 | 10.0 |
| Placement (0-4) | 5848 | 1.784 | 1.455 | 0.0 | 2.0 | 4.0 |

*Panel B. Correlations among the five measured input blocks*

| | SFT predicted idea quality | Off-shelf LLM text score | Execution quality | Connection index (Anderson) | Author ability index (Anderson) |
|---|---|---|---|---|---|
| SFT predicted idea quality | 1.0 | 0.441 | 0.358 | 0.061 | 0.042 |
| Off-shelf LLM text score | 0.441 | 1.0 | 0.107 | 0.081 | 0.031 |
| Execution quality | 0.358 | 0.107 | 1.0 | 0.064 | -0.134 |
| Connection index (Anderson) | 0.061 | 0.081 | 0.064 | 1.0 | 0.155 |
| Author ability index (Anderson) | 0.042 | 0.031 | -0.134 | 0.155 | 1.0 |

## The AI evaluator: leakage versus circularity

The evaluator is a language model *fine-tuned on outcomes* (14). Rather than being asked, cold, to grade a paper, it is trained on a large corpus of institutional traces, each pairing a paper's text with the prestige tier where it eventually landed, and learns to predict that tier from text alone, name-blind. The approach combines two recent advances. The first is *instruction tuning*: supervised fine-tuning teaches a model to map text to a desired output (23), a recipe later extended to align models with human-labeled judgments (24). The second is the *LLM-as-evaluator* literature, in which a language model scores the quality of text on a numeric scale, validated against human raters (15, 22). Fine-tuning an evaluator directly on quality outcomes joins the two. What licenses using its grades as our idea-quality measure is external validation: on the held-out four-tier benchmarks of the companion study, the evaluator reaches roughly 70% exact-tier accuracy in economics, against 25% for chance and about 30% for frontier general-purpose reasoning models.[1] It reads content, and reads it well.

The idea-quality evaluator is the methodological enabling input. Because it was trained on terminal institutional placement traces, its output is a text-legible, pre-publication prediction of quality, the part of a paper's contribution an outcome-trained model can read off the pitch, not a measure of latent intrinsic merit. Because it sees text, not names, it is not mechanically a connection variable. Its limit is that its training target is the journal tier, which makes "regress placement on the score" partly circular.

Two distinct problems must be separated. Leakage is that some scored papers were in the evaluator's training corpus; the overlap is 42 of 6,208 papers (0.68%), and removing them attenuates the idea coefficient by 3.6%. Circularity is deeper and survives even with zero leakage, because the training objective is the journal tier. The empirical strategy is built around this limit, disciplining the use of the score three ways: we report what survives when q is replaced by the off-the-shelf model L, never trained on outcomes (the L-substituted favoritism check appears in Fig. 3b); we difference within authors; and we bound the rest with the identification checks below. The estimand throughout is conditional and bounded: whether connections and prestige predict placement after conditioning on the text-legible quality the model can see and on what an execution rubric can score. It is never "conditional on intrinsic idea quality."

[1] This evaluator is a two-model ensemble fine-tuned to predict a paper's eventual journal tier from its text alone, name-blind at inference (14); the identical ensemble, together with the six-dimension execution rubric used here, is applied and independently validated on a separate corpus of AI- and human-authored economics papers in a companion study (25). The accuracy reported here is exact-tier agreement on a held-out four-tier benchmark, a strict criterion, since adjacent-tier agreement is far higher, and the signal is best calibrated at the apex (SI Table S1). Because its fine-tuning corpus is drawn from the most recent five years of publication outcomes, its temporal overlap with the 2010–2015 cohort studied here is minimal: the direct training-set overlap is 0.68% of the sample, and removing it barely moves the estimates (S4).

## Estimating equation and the three estimands

The estimating equation is a linear five-input regression for paper $i$:

$$Y_i = \alpha + \beta_q q_i + \beta_x X_i + \beta_c C_i + \gamma A_i + \delta L_i + \mu_{f(i)} + \tau_{t(i)} + \theta \,\text{team}_i + \varepsilon_i ,$$

with field fixed effects $\mu_f$, issue-year fixed effects $\tau_t$, team size, and primary-author-clustered standard errors. The headline outcome is prestige placement; the same right-hand side is estimated for Top-5, any publication, the 0–4 ladder, and log citations. Continuous regressors are standardized; binary indicators stay in natural units. The additive form is an empirical claim, not a convenience: the q × C interaction is estimated and found small and insignificant ($\beta = -0.007$, $p = 0.20$ for prestige), so the function is not improved by complementarity. The latent placement-index model underlying this equation, and the meritocracy-versus-network restrictions it casts as testable benchmarks, are derived in SI.

We report three estimands because they answer different questions. Regression coefficients give marginal associations conditional on the other inputs. Shapley decompositions allocate explained $R^2$ across inputs in an order-invariant way, the appropriate tool for "factor shares," since the inputs are correlated and any sequential variance attribution would be arbitrary. Binned placement tables describe the elevator mechanism by comparing placement rates across idea, connection, and execution tiers. Slope, share, and local mechanism are distinct objects, and the analysis requires all three. These map to four estimable tests that each discriminate between the benchmarks: factor shares (is the connection share small and flat, or large and largest at the top?); marginal returns (which input bites at which margin?); elevator structure (is the connection gradient local within an idea tier, or global across tiers, and is it additive or multiplicative?); and apex friction (do the highest-scored papers nonetheless go unmatched?).

## Identification

We do not claim that connections are randomly assigned. Instead, we state one identifying assumption and frame every robustness check as a bound on a named threat to it. The identifying assumption is that, conditional on text-legible idea quality, the off-the-shelf text score, execution, author ability, field, and issue year, the residual connection gradient reflects gatekeeping rather than unobserved paper quality correlated with both connections and placement. The connection estimate is thus a bounded conditional gradient, not a randomized treatment effect.

Five bounding checks address the most plausible violations: omitted text-legible quality (adding the inputs raises, rather than lowers, the idea coefficient); construct circularity of the idea score (addressed by the L substitution, within-author differencing, and the leakage bound above); mechanical name-matching and reverse channels in the connection measure (an acknowledgment-length adjustment and a future-editor placebo, which prior editorship fails and acknowledged-editor links pass); selection into the preprint frame (a Heckman correction that, if anything, raises the connection coefficient); and selection-on-observables fragility (matching, balancing, and identity-resolution variants that move the estimate by single-digit percentages). A separate rotation check, a tier-level difference-in-differences on editor rotations, corroborates a stock-of-connection effect that survives quality controls but cannot certify a discrete jump at editor entry, so we do not lean on it for identification. The threat-by-threat detail, all robustness numbers, and the broad-picture robustness (alternative outcomes, the 216-specification curve, and cohort splits) are reported in SI.

## Results

We read the placement data through the three lenses defined in Methods: marginal coefficients, order-invariant Shapley shares, and binned placement tables.

### Factor shares: a sequence, not a contest

Figure 1 reports the production-function coefficients with 95% confidence intervals, and an order-invariant Shapley decomposition partitions the explained variance in prestige placement (top-field journal or better) into the five input blocks. The full multi-indicator model explains 20.6% of the variance in prestige placement. The Shapley partition (summing to 0.196, with the residual 0.010 carried by field, year, and team-size controls outside the five input blocks) ranks the inputs: execution 37.3%, connections 24.2%, text-legible idea quality 16.4%, author ability 13.9%, and the off-the-shelf model score 8.2% (Fig. 2; SI Table B1). Execution, meaning how rigorously a paper is written, identified, and defended, is the single largest input.

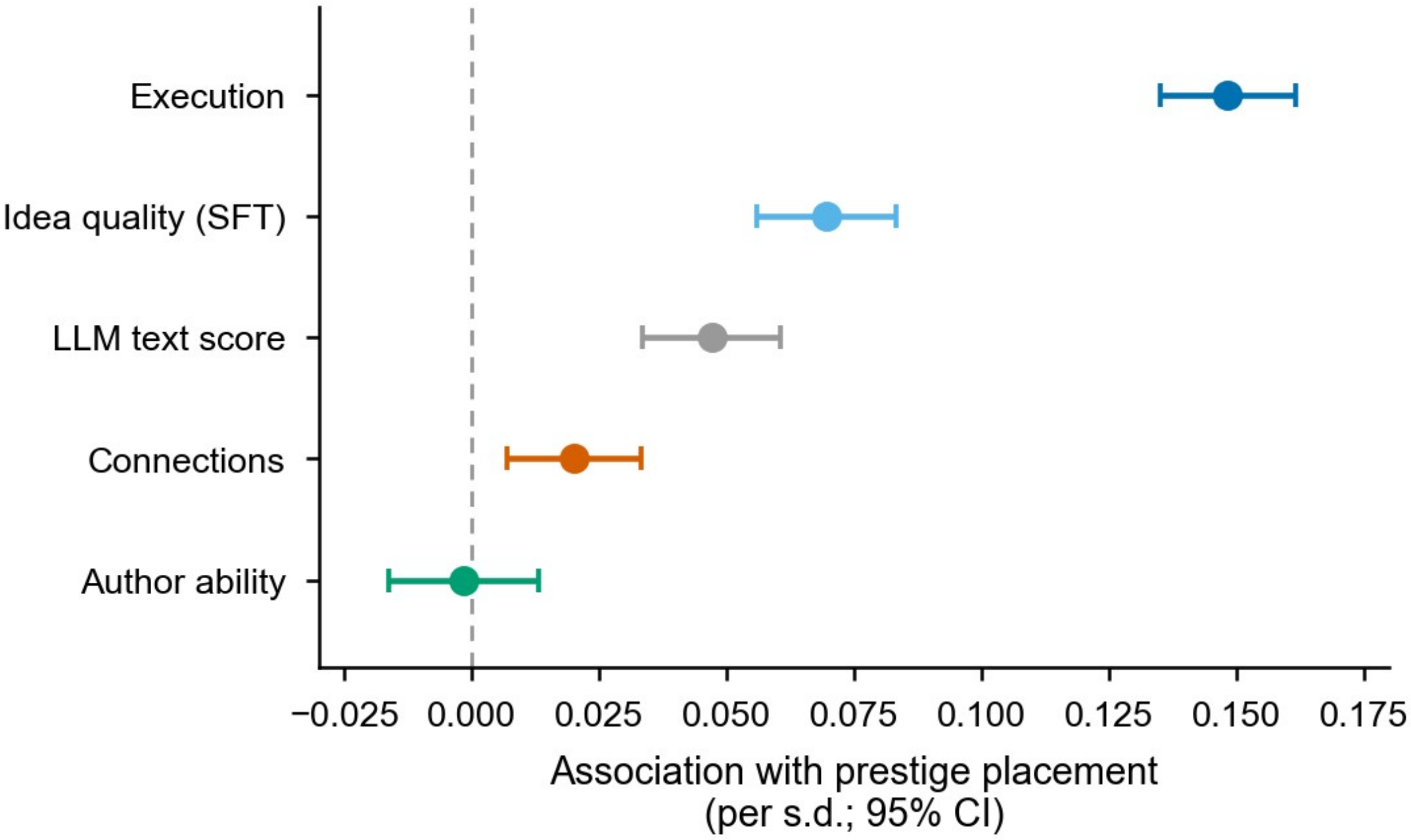

**Figure 1 | Production-function coefficients for prestige placement.**

Points are standardized OLS coefficients (compact five-input specification); horizontal bars are 95% confidence intervals from primary-author-clustered standard errors (n = 5,848). The dashed line marks zero. Execution is the largest input; author ability is indistinguishable from zero once the other inputs are present.

But the ranking depends on where on the ladder one stands. Figure 2 plots the Shapley shares across the five placement margins. The connection share rises monotonically with selectivity, from 15.3% at the any-publication margin, to 18.7% at mid-tier-or-better, 19.2% at top-field, 24.2% at prestige, and 38.8% at the most selective margin, placement in one of the five most prestigious journals (the "Top-5"). The execution share moves inversely, falling from 61.5% at any-publication down to 15.6% at the Top-5 margin. This is the cleanest statement of the sequence: the question "will this be published at all?" is answered mostly by execution; the question "will this reach the apex journals?" is answered substantially by connections; and text-legible idea quality grades the rungs in between. Gatekeeping is real and concentrated precisely at the apex, where a network account expects it. Yet across the function as a whole it is not the largest input, as a meritocratic account expects of the broad margins. Both accounts are locally right and globally incomplete.

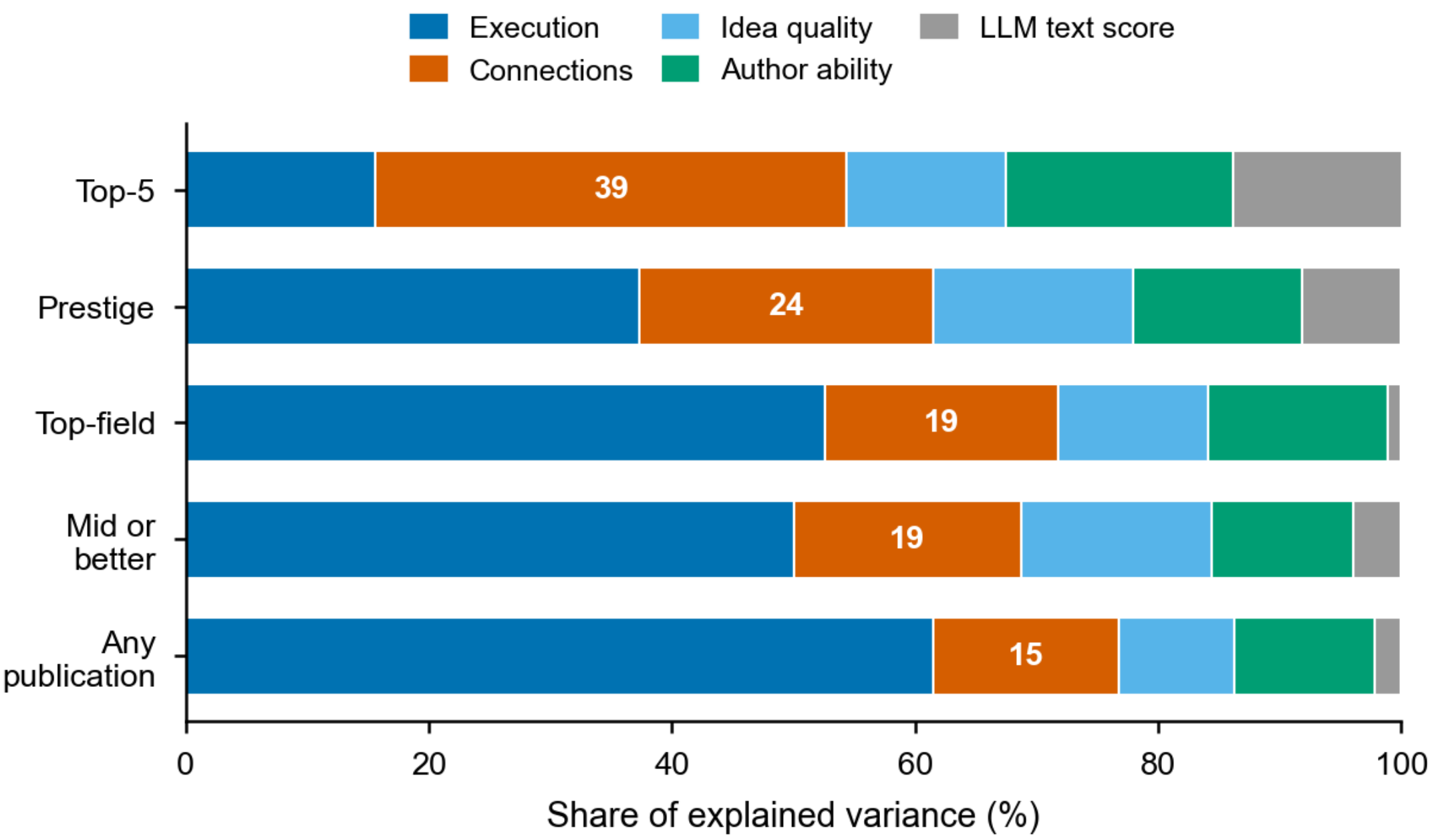

**Figure 2 | Factor shares across placement margins.**
Each bar partitions the explained variance at one outcome margin into order-invariant Shapley shares. The connection share (orange, labelled) rises monotonically from 15% at any-publication to 39% at the Top-5 apex, while the execution share falls.

In slope form (Fig. 1), the compact index specification gives execution the largest coefficient of the five inputs (+0.148), then text-legible idea quality (+0.070), the off-the-shelf score (+0.047), and the connection index (+0.020); the author-ability index is essentially zero (−0.002, not significant) once the other blocks are present, indicating that author ability operates through the other channels rather than as a freestanding input. In the multi-indicator model, the acknowledged-editor coefficient is +0.051 (SI Table S2). These are conditional associations with all other blocks held fixed.

The two readings are complementary, not contradictory: a coefficient is a per-standard-deviation return, while a factor share is the unique, multi-variable variance a block accounts for, and the two can rank inputs differently. The connection share in Figure 2 exceeds the idea-quality share even though the connection index carries the smaller coefficient in Figure 1, for two reasons. First, the share credits the full multi-indicator connection block (acknowledged editors, editor tenure, coauthor and advisor proximity, institutional prestige, and network centrality), whose first principal component explains only about 31% of its variance, so a single scalar index compresses it and understates its slope. Second, connections are nearly orthogonal to the quality signals (correlations about 0.06), so their explained variance is almost entirely unique, whereas idea quality is correlated with execution and the off-the-shelf score (about 0.36 to 0.44) and therefore shares its variance credit with them. Per standard deviation, a unit of idea quality moves placement more than a unit of connection; across the whole population the connection block nonetheless accounts for more of the explained variation, and most sharply at the apex.

## The dual channel: capture and favoritism, additive and bounded

This is the central result. Connections are associated with better placement through two channels that prior work has conflated, and separating them dissolves the merit-versus-network debate. Figure 3 shows the favoritism channel directly; the full dual-channel ledger (capture, favoritism, additivity, and the off-the-shelf-score replication) is in SI Table S3.

*Capture.* Connected authors write papers that score higher on the text signals, but the premium is a discrete step, not a smooth gradient. The mean idea-quality score rises from 2.89 with no connection to 3.19 at low connection, then is essentially flat across the higher tiers (3.14 at mid, 3.22 at high); execution behaves identically (3.50, then 3.91, 3.92, 3.93). Capture is an any-connection premium of roughly 0.3–0.4 score points, about 0.6 standard deviations on the idea signal between the unconnected and the connected, not a return that grows with connection intensity. This is why the linear correlation between the connection index and the quality signals is near zero (0.06 in Table 1) even though the tier means differ: connected authors submit better-scoring work, but among the connected, more connection does not buy higher scores. Some of what looks like favoritism in any single-channel regression is this any-connection step. The wording matters for the estimand: high-connection papers score higher on the text-legible signals, which is not a claim that connections produce better ideas.

*Favoritism.* Holding the idea signal fixed, connected papers still place better. Within fair-rated papers, equal on the ex-ante idea tier, the Top-5 rate rises from 1.9% with no connection to 14.1% with high connection, a 7.2-fold gap that equal text-legible quality cannot explain. The same within-tier gradient appears for strong-rated (7.0% to 22.7%, 3.3-fold) and exceptional-rated (16.8% to 39.5%, 2.4-fold) papers (Fig. 3a). The gradient is steepest where there is most room to move and shallowest at the top, but it is everywhere positive and everywhere within idea tier.

*Additivity.* The two channels add; they do not multiply. The quality-by-connection interaction is economically negligible and statistically insignificant: $\beta = -0.007$ ($p = 0.20$) for prestige placement, $\beta = +0.007$ ($p = 0.11$) for the Top-5 margin. Connections do not "unlock" the value of good ideas, as a strong network account requires; they shift placement by a roughly constant amount on top of whatever the idea signal delivers. Capture and favoritism are separable, additive forces within one specification, not two readings of a single multiplicative network effect, consistent with the near-zero capture correlation in Table 1.

The favoritism gradient is not an artifact of training the score on the outcome. The concern is that q is trained on the journal tier, so a within-q favoritism gradient could in principle reflect circularity rather than gatekeeping. We therefore repeat the favoritism test using the off-the-shelf model L, which was never trained on placement, in place of q (Fig. 3b). Holding L fixed in terciles, the connection gradient survives intact: within the lowest-L tercile the Top-5 rate rises from 3.7% with no connection to 16.7% with high connection (4.6-fold), and the gradient persists at every L tercile (Mid-L 6.5% to 19.6%; High-L 12.9% to 36.7%). Because L carries no outcome-training circularity, the dual channel remains visible through a quality lens that the circularity critique cannot touch.

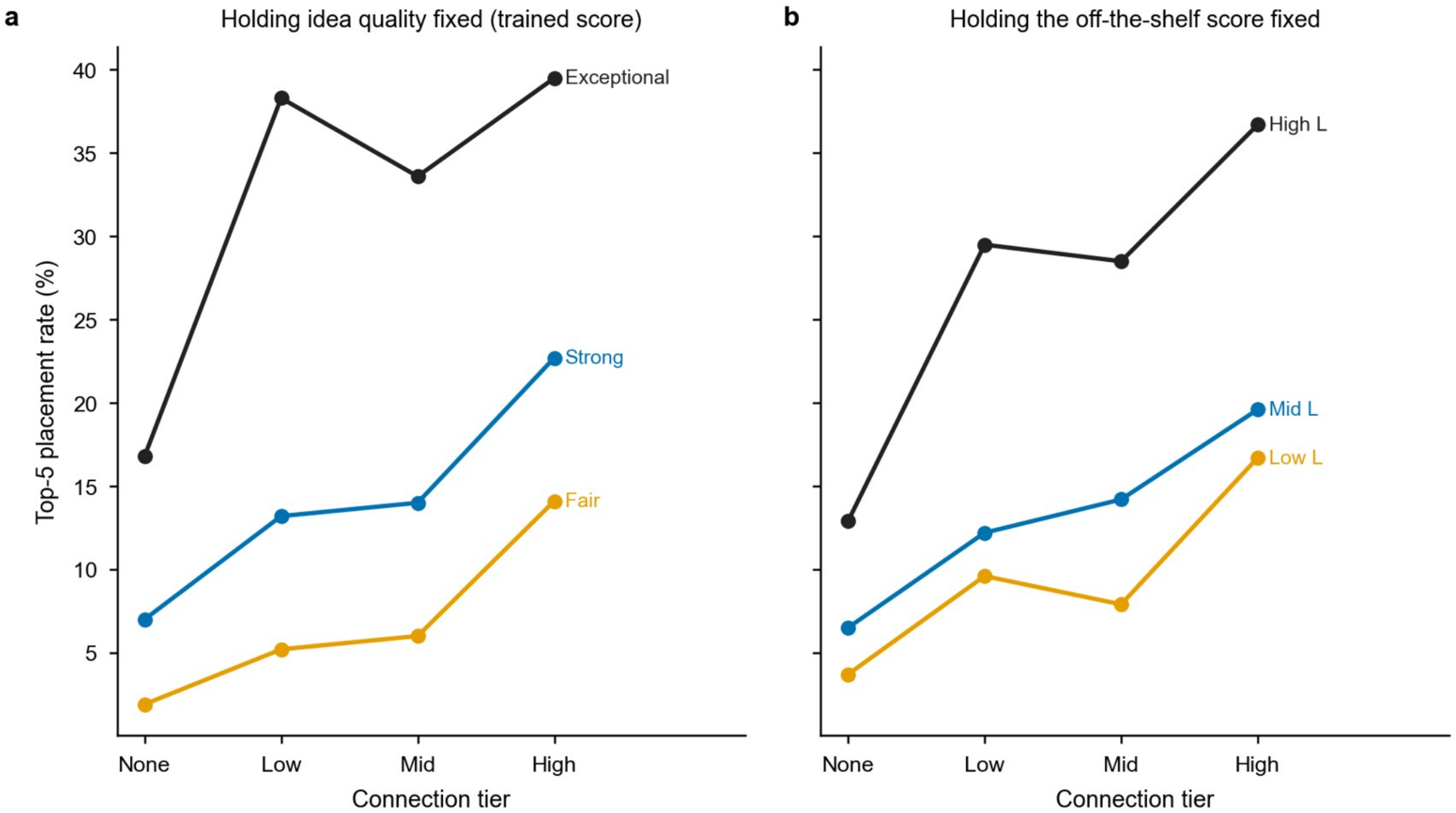

**Figure 3 | The dual channel: favoritism survives holding quality fixed.**
Top-5 placement rate by connection tier, within tiers of (a) the trained idea-quality score and (b) the off-the-shelf language-model score, which was never trained on the placement outcome. The connection gradient persists in both panels, so the favoritism channel is not an artifact of training the score on placement.

The favoritism channel is bounded. Connections shift a paper's placement distribution up the ladder, but where the mass lands depends on the idea tier, and the balance between the apex and the next rung down tips toward the apex only at the top. Two facts make this precise, and they must be read together. First, connections raise both prestige rungs at every idea tier: for fair-rated papers, high connection lifts the top-field rate from 12.8% to 30.3% and the Top-5 rate from 1.9% to 14.1%. So a fair-idea paper can reach the apex (14.1% do at high connection), but top-field, not the Top-5, remains its modal prestige destination even at high connection. Second, the Top-5-minus-top-field rate gap within each tier (SI Table S4) expresses the balance. For exceptional papers the gap turns positive as connections rise (from −0.10 to +0.17): the modal prestige destination flips from top-field to the apex. For strong and fair papers the gap stays negative (strong −0.20 to −0.10; fair −0.11 to −0.16): connections move these papers up the ladder, but top-field stays their typical landing spot. The bounded reading is therefore not that fair papers never reach the apex, since they do, but that connections do not make the apex the typical outcome for ordinary ideas; they raise the odds of every rung and tip the apex into reach mainly for ideas the signal already ranks at the top. So "it's who you know" and "it's the work" are both true: the work sets where the placement distribution is centered, connections shift it up by about one rung, and only strong work plus connections centers it on the apex.

## Execution sets the meritocratic floor

Execution plays a different role from connections, as a comparison of margins shows (SI Table S5). Among low-idea-quality papers, strong rather than weak execution is associated with a top-field-or-better rate of 48.1% versus 11.8%, and it cuts the unmatched rate (the share never matched to any of the four journal tiers) from 46.9% to 20.6%. Among high-idea-quality papers, weak execution leaves 40.1% unmatched, while strong execution lifts the Top-5 rate to 35.7%. Execution moves both the floor (publish-or-not) and the ceiling (the apex).

Connections do the opposite. Across connection tiers the unmatched rate is essentially flat (about 32% with no connection and 27% with high connection); connections add almost nothing at the publish-or-not margin and almost everything at the apex margin. Execution is the meritocratic floor: without it, even a good idea falls through. Connections are the favoritism ceiling, biting only near the apex. A favoritism regression that omits execution therefore attributes to connections variance that belongs to the largest, and most floor-relevant, input.

### Marginal returns: placement and citations reward different inputs

The inputs that win placement are not the inputs that win citations (26), and the contrast sharpens what the idea signal measures. In the cross-outcome panel (field and issue-year fixed effects), the execution coefficient on log citations is +0.602 per standard deviation, roughly five times the idea-quality coefficient of +0.123; the off-the-shelf score is insignificant and wrong-signed. (Within realized journal tier, a stricter specification, the idea-quality coefficient on citations is a smaller but correctly signed +0.054, significant at the 5% level; the cross-panel and within-tier numbers are kept labeled separately because they answer different questions.) Execution out-predicts the idea signal on downstream impact by about fivefold, while the idea signal is the stronger predictor of placement at the top. This is the calibration the design promised: the tier-trained signal captures "what the top journals take," not "what gets cited," and execution captures more of the latter. Favoritism regressions that omit execution have been measuring the weaker predictor of impact.

### Apex friction: the signal predicts the ceiling, not the floor

Figure 4 and SI Table A1 quantify the friction that remains. Among the top 1% of papers by the idea-quality score, 57.1% reach the apex journals and 17.5% are unmatched to the four-tier journal panel; among the bottom half of the score distribution, 5.8% reach the apex and 34.9% are unmatched. The signal generates a tenfold spread in apex rates (5.8% to 57.1%) but only a twofold spread in unmatched rates (34.9% to 17.5%). It sharply predicts the ceiling and only weakly protects against the floor. The top-percentile cut contains 63 papers, so we treat the 17.5% as a headline diagnostic and report the top-5% and top-10% cuts (21.9% and 25.3% unmatched) in the appendix rather than over-reading finer sub-cuts. The qualitative fact is robust across cuts: even the papers the discipline-trained signal ranks highest face nontrivial publication-market friction. This is not evidence that those papers were truly excellent; it is evidence that high text-legible quality is not a sufficient condition for reaching the visible journal ladder (27).

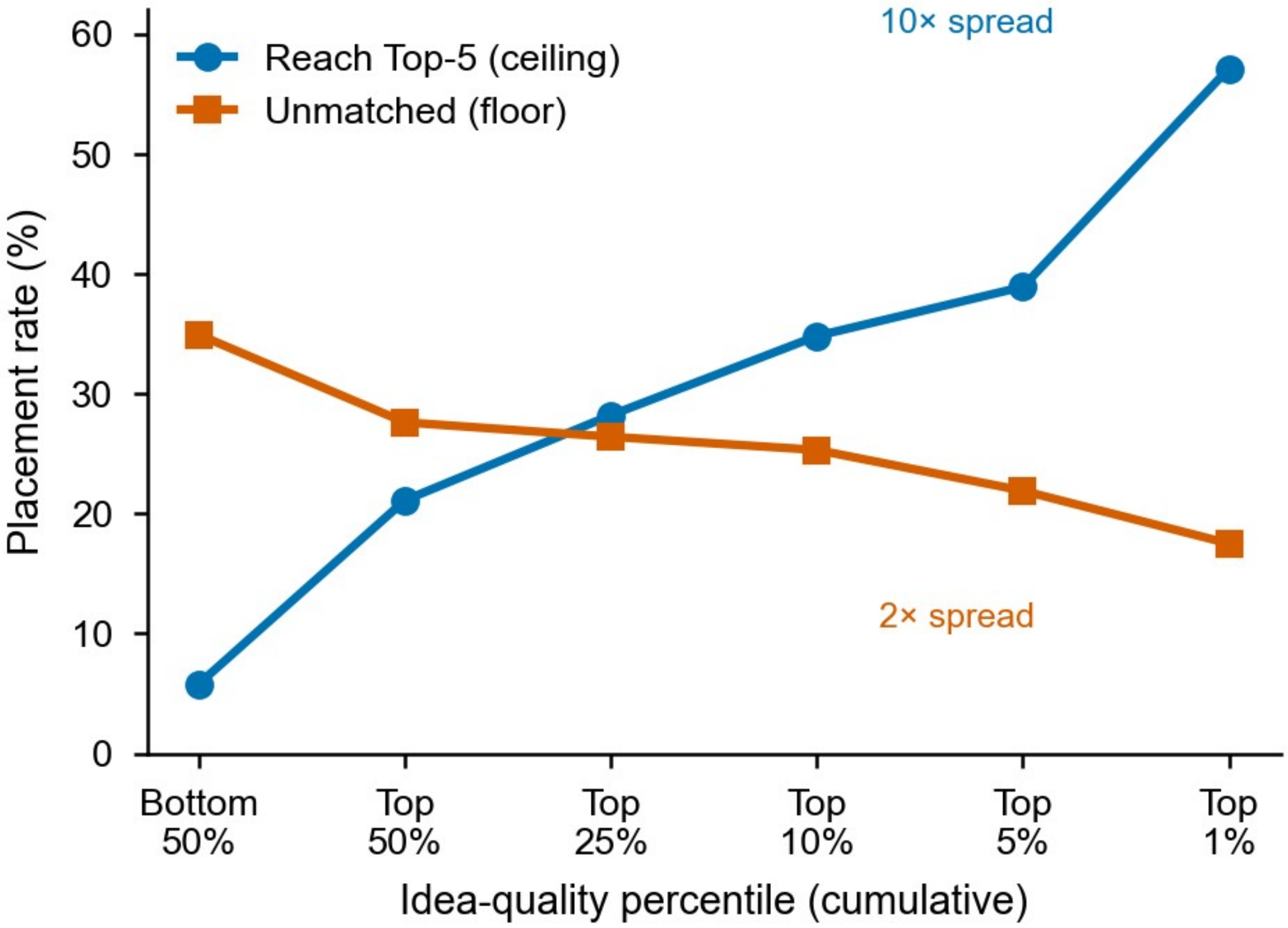

**Figure 4 | Apex friction.**

Share of papers reaching the Top-5 (ceiling) and going unmatched to the journal panel (floor) by cumulative idea-quality percentile. The signal drives a roughly tenfold spread in the ceiling but only a twofold spread in the floor: high text-legible quality buys a shot at the top, not insurance against the floor.

## Discussion

Our findings do not choose between the meritocracy view of scientific publishing and the network view; they size both and show that they describe different channels of one machine. Authors with editorial connections write papers that score higher on the observed text and execution measures. This is capture: advantage that lives in the inputs. Holding those measures fixed, connected papers still place higher near the top of the prestige ladder. This is favoritism: advantage that lives in the placement itself. The two enter the same placement production function additively rather than multiplicatively, and the favoritism component is bounded: it raises the odds of every rung but makes the apex the typical destination only for the most highly rated ideas. The reader who believes publishing is meritocratic and the reader who believes it is networked are each describing one channel of a two-channel system; the data require both and rank them by where on the ladder they bite. So the colloquial conviction that “it’s who you know” and the rejoinder that “it’s the work” are simultaneously correct. The work sets where a paper’s placement distribution is centered, connections shift it up, and only strong work combined with connections centers it on the apex.

The sharpest implication for prior work is methodological, and it generalizes beyond economics. Because execution quality, the rigor and robustness with which an idea is carried out, is the single largest input and is itself correlated with connections, any analysis that estimates a favoritism gradient without controlling for execution is a proxy regression in which the connection term silently absorbs unmeasured execution. This is, to our knowledge, the first study to measure execution quality from the full text before the publication outcome at this scale and net it out, and doing so is precisely what licenses reading the residual favoritism gradient as gatekeeping rather than as mismeasured rigor. The point is not that earlier estimates were wrong but that they bundled together two inputs the present method can now separate. The caution applies to any audit of prestige allocation in which a hard-to-observe quality-of-execution variable is correlated with the social-network variable under study.

The structure we document, a connection premium that is bounded to roughly one rung, additive rather than multiplicative, and concentrated at the apex, is informative about why editors place connected papers higher, even though it does not fully resolve the question. Two mechanisms can each generate a positive within-tier connection gradient. Under pure relationship rents, gatekeepers trade quality for relationships, and connected papers occupy slots that better unconnected work would otherwise have won; this is inefficient. Under soft information, connections convey a signal about quality that the written pitch (and the referee) cannot fully see, such as a credible revision plan or a track record of delivering on a research agenda, and gatekeepers who weight it are improving, not degrading, selection. Our data tilt against the pure-rent reading on two counts. First, the gradient is bounded in where it lands: connections raise the apex rate of fair-rated papers (the Top-5 rate rises to 14.1 percent) but lift them mostly into the tier just below the apex (the top-field rate rises to 30.3 percent), so the apex does not become the typical destination for ordinary ideas; the balance tips into the apex only for ideas the signal already ranks at the top. Wholesale slot-capture by low-quality connected work would require exactly the opposite. Pure rents need not respect the idea tier; a soft-information nudge naturally tracks it. Second, the connection channel is not landing low-impact papers at high-prestige venues: acknowledged-editor links carry a positive and significant association with downstream citations in the cross-outcome panel (about +0.16 per standard-deviation-equivalent), so connected papers are, if anything, cited more, not less, than their placement would predict. That is hard to reconcile with gatekeepers substituting relationships for quality and easy to reconcile with connections conveying quality the text signal misses. We stop short of declaring the gradient efficient, because soft information and rents can coexist and we cannot price the welfare loss to the marginal displaced paper. But the anatomy is more consistent with a bounded informational nudge than with apex slot capture.

We cannot price welfare here. We do not observe every rejected submission, every private revision path, or every reason a working paper goes unpublished. What we can document is a residual friction in a high-prestige preprint universe: a nontrivial share of papers the text signal ranks at the very top do not reach the visible journal ladder at all, and the connection and execution measures help account for who lands where. These data do not settle whether the favoritism gradient is inefficient or efficient. What they do settle is the anatomy: where each input bites, how large each one is, and that the favoritism channel is real, bounded, and concentrated at the apex.

The vocabulary this study introduces, factor shares for the inputs to prestige, marginal returns that differ by margin, a bounded elevator for connections, and apex friction for the best ideas, travels to any prestige market that has text-legible inputs and a measurable outcome ladder. Grant funding, faculty hiring (28), and conference acceptance are natural next cases: each allocates scarce prestige, each has an ex-ante written record (a proposal, a job-market paper, a submission) that a name-blind evaluator could score before the decision, and in each the merit-versus-networks debate has been stuck for the same reason it has been stuck in publishing: there was no quality measure free of the very network one wants to net out. The method is portable; the estimates are not. The papers studied here are an already-connected, already-selected elite population, and the apex journals are unusual institutions; this is the economics publication game among a specific affiliated cohort, not a universal law of science. Two limits are central. The AI idea-quality score is trained on terminal institutional placement traces. This is simultaneously its value (it reads text, not names) and its weakness (its training target is the outcome we then study), which is why an off-the-shelf model never trained on those traces, together with within-author comparisons, carries the identification weight rather than the trained score alone. And the connection measures drawn from public records are incomplete: acknowledgments reveal only some contact, editorial-board panels cover only some venues and years, and author disambiguation is imperfect. Connections here are observed, not randomly assigned; our claims about favoritism are associations that survive a stack of adjustments, not the effect of a randomized intervention.

For three decades the question of whether good ideas or good connections decide where scientific work lands resisted answer, because quality could not be measured before the outcome without using the outcome as the yardstick. An ex-ante, name-blind, text-derived measure of idea quality removes that constraint, and with it we estimate a five-input production function for journal placement across a large cohort of economics working papers. The answer is not “ideas” or “connections” but a structure. Execution sets a meritocratic floor and is the largest input on the broad margins; text-legible idea quality grades the ladder; and connections set a favoritism ceiling that bites near the apex through two additive channels: connected authors write higher-scoring papers, and at equal scores they still place about one rung higher. Even the highest-scoring papers face real friction reaching the visible ladder. The publication game is a meritocratic floor with a favoritism ceiling, and most of the work in between is done by execution. The method that delivers this, an auditable text-derived quality measure used with explicit discipline about its training target, is portable to the other prestige markets that science runs on, and the accounting it enables is, we hope, a more honest vocabulary for a debate that has too long been forced into a false choice.

# Supplementary Information for “Merit or networks? What decides where research is published”

Numeric citations (N) refer to the reference list in the main text.

This section collects the technical detail underlying the Methods and Results: the latent placement-index model that casts meritocracy and network as testable restrictions (S1), the step-by-step sample construction and the multinomial rationale for the prestige binary (S2), the full per-dimension execution validation (S3), the threat-by-threat identification stack with all numbers (S4), the broad-picture robustness across outcomes, specifications, and cohorts (S5), the heterogeneity of the dual channel across fields, teams, and time (S6), and data availability (S7).

## S1. The latent placement-index model and its meritocracy-versus-network restrictions

The framework casts the two competing accounts of publishing (meritocracy and network) as restrictions on a latent placement-index model, so that the empirical objects in the main text are tests rather than descriptions. Paper *i* has a latent placement index

$$Y_i^* = \alpha + \beta_q q_i + \beta_x X_i + \beta_c C_i + \lambda(q_i \times C_i) + \gamma A_i + \delta L_i + \varepsilon_i,$$

and observed placement is an ordered coarsening of $Y_i^*$ into the five-rung ladder with tier-specific thresholds $\{\kappa_r\}$, so paper *i* reaches rung *r* when $Y_i^* \geq \kappa_r$. The meritocratic benchmark is separability with no residual gatekeeping: conditional on the legible inputs, $\beta_c = 0$ and $\lambda = 0$, so the probability of clearing any threshold is independent of connections, and the highest-q papers clear the top threshold regardless of C. The discretionary-network alternative is complementarity with an unbounded “elevator”: $\beta_c \gg 0$ and $\lambda > 0$, with the connection shift large enough relative to the top spacing $\kappa_4 - \kappa_3$ that even low-q connected papers clear the apex. The reading the data support is the intermediate restriction $\beta_c > 0$, $\lambda \approx 0$, with a connection shift comparable to one threshold spacing and roughly constant across rungs: connections move the index additively by about one rung, so a paper’s idea tier sets which thresholds are in reach and connections determine which of those it clears. Capture (the dependence of the inputs themselves on connections, $E[q / C]$ increasing) is then a separate object that can coexist with $\lambda \approx 0$: capture lives in the inputs, favoritism in the index, and additivity is what keeps them separable.

The estimating equation reported in Methods is the linear analog of this index, estimated on paper *i*:

$$Y_i = \alpha + \beta_q q_i + \beta_x X_i + \beta_c C_i + \gamma A_i + \delta L_i + \mu_{f(i)} + \tau_{t(i)} + \theta\,\text{team}_i + \varepsilon_i,$$

with field fixed effects $\mu_f$, issue-year fixed effects $\tau_t$, team size, and primary-author-clustered standard errors. The additive form is an empirical claim: the q × C interaction is estimated and found small and insignificant (β = −0.007, p = 0.20 for prestige), so complementarity does not improve the function. Where binned descriptive tables show larger absolute gradients at high q, we report them as descriptive heterogeneity, not as a rejection of additivity. The three estimands of Methods map to four estimable tests that each discriminate between the benchmarks: factor shares (is the connection share small and flat, or large and largest at the top?); marginal returns (which input bites at which margin?); elevator structure (is the connection gradient local within an idea tier, or global across tiers, and is it additive or multiplicative?); and apex friction (do the highest-scored papers nonetheless go unmatched?).

Table S2. Production-function regressions: compact-index and multi-indicator specifications (standardized continuous regressors; primary-author-clustered SEs in parentheses). Visualized in Figure 1.

| Predictor | Compact index specification | Multi-indicator benchmark |
|---|---|---|
| SFT predicted idea quality | +0.0695 (0.0070) | +0.0489 (0.0071) |
| Off-shelf LLM text score | +0.0471 (0.0069) | +0.0377 (0.0068) |
| Execution quality | +0.1483 (0.0068) | +0.1249 (0.0071) |

| Predictor | Compact index specification | Multi-indicator benchmark |
|---|---|---|
| Connection index (Anderson) | +0.0201 (0.0067) | |
| Author ability index (Anderson) | -0.0015 (0.0075) | |
| Acknowledged editors | | +0.0511 (0.0069) |
| Additional connection indicators | Index only | Included |
| Additional author-ability indicators | Index only | Included |
| Team size, field FE, issue-year FE | Included | Included |
| N | 5848 | 5848 |
| R2 | 0.1746 | 0.2062 |

## S2. Sample construction and the multinomial rationale for the prestige binary

SI Table A0 reports the step-by-step construction from the 6,210 issued NBER working papers to the 6,208 scored papers and the 5,848-paper complete-case production-function sample (3,067 primary-author clusters). Measurement and calibration exhibits use all 6,208 scored papers, because calibrating the idea-quality signal against realized placement needs neither the connection nor the ability data; the five-input estimates use the 5,848-paper complete-case sample. Every reported number is tagged to one of these two bases, and the two are never mixed. The substantive conclusions are insensitive to the choice of base: calibration and percentile diagnostics computed on the 5,848-paper sample differ from the 6,208-paper versions only in the third significant figure.

Table A0. Sample construction.

| Sample step | Papers | Notes |
|---|---|---|
| Analysis dataset | 6210 | One row per NBER working paper in the joined analysis dataset |
| Scored on SFT predicted idea quality | 6208 | Papers with nonmissing SFT score |
| Nonmissing SFT, LLM, and execution | 6178 | Before ability/connection complete-case restrictions |
| Headline five-block analytical sample | 5848 | Complete case on five blocks; 3,067 primary-author clusters |
| DV-B prestige placements in headline sample | 2325 | placement ≥ top-field in the 5,848-paper sample |

Table A1. Placement distribution by idea-quality percentile. Visualized in Figure 4.

| q cut | N | Top-5 | Top-field | Mid | Lower | Unmatched | Mean placement |
|---|---|---|---|---|---|---|---|
| top 1% | 63 | 57.1 | 11.1 | 9.5 | 4.8 | 17.5 | 2.86 |
| top 5% | 311 | 38.9 | 22.8 | 9.3 | 7.1 | 21.9 | 2.50 |
| top 10% | 621 | 34.8 | 24.2 | 9.8 | 6.0 | 25.3 | 2.37 |
| top 25% | 1552 | 28.2 | 28.2 | 11.0 | 6.3 | 26.4 | 2.25 |
| top 50% | 3104 | 21.1 | 30.3 | 12.7 | 8.2 | 27.6 | 2.09 |
| bottom 50% | 3104 | 5.8 | 22.0 | 19.7 | 17.7 | 34.9 | 1.46 |

Privileging the prestige binary is a substantive choice. A multinomial decomposition shows that the idea-quality signal is monotone and strongest at the top of the ladder and can be negatively associated with lower-tier placement relative to remaining unmatched: the signal orders the apex cleanly but does not order "lower-tier versus unpublished." The prestige binary is therefore the most interpretable gatekeeping margin, and the conclusions hold across the alternative outcomes.

On calibration, beyond the headline of SI Table S1: among papers that realized Top-5 placement, 53.3% carry an ensemble argmax grade of "exceptional" and 94.1% of "strong or exceptional," against a mean score of 3.36 for Top-5 papers versus 2.85 for unmatched papers; the dangerous-error rate is small, with only 0.7% of Top-5 papers ensemble-rated "limited." We report this as the ensemble-argmax calibration; a best-of-two-models rule yields higher headline percentages and is not used, to avoid presenting two numbers for one concept.

Table S1. Idea-quality calibration by realized placement (6,208 scored papers; ensemble-argmax shares).

| Placement | N | Exceptional (%) | Strong or exceptional (%) | Limited (%) | Mean q |
|---|---|---|---|---|---|
| unmatched | 1940 | 23.6% | 69.5% | 3.1% | 2.854 |
| lower | 804 | 15.7% | 62.2% | 4.5% | 2.687 |
| mid | 1006 | 19.9% | 72.3% | 2.6% | 2.845 |

| Placement | N | Exceptional (%) | Strong or exceptional (%) | Limited (%) | Mean q |
|---|---|---|---|---|---|
| top-field | 1622 | 28.8% | 86.1% | 0.7% | 3.080 |
| Top-5 | 836 | 53.3% | 94.1% | 0.7% | 3.361 |

## S3. Per-dimension execution validation

The execution rubric scores six dimensions, produced independently of the idea score: writing and presentation, identification, econometric methodology, mechanism and external validity, contribution, and robustness. The composite is close to unidimensional: the first principal component explains 68.6% of the across-dimension variance. Per-dimension regressions are nonetheless retained because the writing-and-presentation dimension carries surprising weight, and "what counts as good execution" is field-specific: the execution dimension most correlated with prestige is writing and presentation in the theory-leaning fields and identification strategy in the applied empirical fields, even though execution is the largest input everywhere.

Table S5. Execution gradients: placement rates by idea-quality tier and execution tier.

| q tier | Execution tier | N | P(Top-5) | P(top-field+) | P(unmatched) | Mean placement |
|---|---|---|---|---|---|---|
| LOW | LOW | 1062 | 3.2 | 11.8 | 46.9 | 0.98 |
| LOW | MID | 597 | 3.4 | 32.3 | 28.3 | 1.62 |
| LOW | HIGH | 291 | 8.2 | 48.1 | 20.6 | 2.05 |
| MID | LOW | 541 | 6.1 | 23.5 | 39.4 | 1.32 |
| MID | MID | 751 | 9.3 | 41.1 | 25.0 | 1.89 |
| MID | HIGH | 657 | 19.0 | 56.0 | 24.0 | 2.19 |
| HIGH | LOW | 394 | 9.6 | 32.0 | 40.1 | 1.49 |
| HIGH | MID | 628 | 16.9 | 50.2 | 26.4 | 2.05 |
| HIGH | HIGH | 927 | 35.7 | 67.1 | 21.0 | 2.56 |

The off-the-shelf score L is free of outcome-training circularity but not, by itself, of all prestige bias: off-the-shelf models carry pretrained priors and, given identifying cues, can rate prestigious-looking work more highly (17). Scoring a short, name-blind pitch (no author, institution, or venue strings at inference) limits those priors here, and the within-author check (S4) bounds what remains.

## S4. Identification: the threat-by-threat stack

The identifying assumption is stated in Methods; the checks below bound the most plausible violations of it, drawing on several estimators rather than one. The connection estimate is a bounded conditional gradient, not a randomized treatment effect.

### *Threat 1: omitted text-legible quality*

If connected papers are simply better in ways the regression does not see, the gradient is spurious. The five-block specification addresses the legible part of quality, and adding the inputs raises the idea coefficient: collapsing the multi-indicator connection block to a single index moves the idea coefficient from +0.049 to +0.070, because the multi-indicator block had been absorbing idea-quality variance through the connection–quality correlation. The "execution dominates" and "favoritism survives" results both strengthen, not weaken, under the parameter-parallel specification.

### *Threat 2: construct circularity of the idea score (distinct from training-set leakage)*

Leakage is bounded (42/6,208 papers; 3.6% attenuation on removal). Circularity is addressed three ways. First, the off-the-shelf model L, never trained on outcomes, reproduces the capture and favoritism patterns; the favoritism gradient holding L fixed in terciles is reported in the main text (Fig. 3b; full ledger in SI Table S3), so the dual channel is not an artifact of the tier-trained objective. Second, within-author fixed effects (4,119 papers by 1,174 authors with two or more cohort papers) leave the idea coefficient at +0.183 (from +0.270 pooled): the same author's higher-scored papers place better, so the score is not merely a reputation proxy. Third, the idea signal beats the off-the-shelf model in seven of seven outcomes and is correctly signed in within-tier citation regressions where the off-the-shelf incremental signal is wrong-signed. The idea score is a sharper but bounded proxy, and its value is the variance it explains over the untrained model.

Table S3. Dual-channel ledger. Panel A capture; Panel B favoritism; Panel C additivity; Panel D favoritism under the off-the-shelf score. Panels B and D are visualized in Figure 3.

*A. CAPTURE: mean input score by connection tier*

| Connection tier | SFT idea quality | LLM text score | Execution | N |
|---|---|---|---|---|
| No connection | 2.889 | 2.288 | 3.500 | 2895 |
| Low | 3.188 | 2.390 | 3.911 | 984 |
| Mid | 3.145 | 2.385 | 3.919 | 985 |
| High | 3.224 | 2.435 | 3.926 | 984 |
| Delta (High - None) | +0.335 | +0.147 | +0.426 | |

*B. FAVORITISM: P(Top-5) within idea tier*

| Idea tier | P(Top-5) no connection | P(Top-5) high connection | High/none ratio | N (none / high) |
|---|---|---|---|---|
| Fair | 1.9% | 14.1% | 7.2x | 977 / 142 |
| Strong | 7.0% | 22.7% | 3.3x | 1402 / 541 |
| Exceptional | 16.8% | 39.5% | 2.4x | 369 / 281 |

*C. ADDITIVITY: quality x connection interaction is null*

| Outcome | q x connection | SE | p-value | N |
|---|---|---|---|---|
| Prestige (Top-5 + top-field) | -0.0074 | 0.0058 | 0.203 | 5848 |
| Top-5 | +0.0073 | 0.0046 | 0.111 | 5848 |

*D. FAVORITISM survives the off-the-shelf model (no outcome-training): P(Top-5) within off-the-shelf score (L) tier*

| Off-the-shelf score (L) tier | P(Top-5) no connection | P(Top-5) high connection | High/none ratio | N (none / high) |
|---|---|---|---|---|
| Low-L | 3.7% | 16.7% | 4.6x | 1582 / 365 |
| Mid-L | 6.5% | 19.6% | 3.0x | 647 / 184 |
| High-L | 12.9% | 36.7% | 2.8x | 666 / 436 |

Table S4. Bounded-elevator asymmetry: within-idea-tier shift in P(Top-5) minus P(top-field) from no to high connection.

| Idea tier | None: P(Top5)-P(TF) | High: P(Top5)-P(TF) | Shift (None->High) | Crosses into Top-5? |
|---|---|---|---|---|
| Exceptional | -0.1 | 0.174 | 0.275 | Yes (+) |
| Strong | -0.202 | -0.104 | 0.098 | No (stays in top-field) |
| Fair | -0.108 | -0.162 | -0.053 | No (stays in top-field) |

### *Threat 3: mechanical name-matching and reverse channels in the connection measure*

The acknowledgment channel is the strongest connection measure and the one most mechanically tied to names in the text; two checks bound it. Adding acknowledgment word count attenuates the prestige estimate by 32.3%, a real mechanical component, so we treat the raw acknowledgment coefficient as an upper association and the length-adjusted value as conservative. A future-editor placebo is the central construct-validity audit: prior editorship fails it (future editorship predicts placement at least as strongly as past editorship, indicating talent or career trajectory rather than gatekeeping, so that channel is demoted), while acknowledged-editor links pass. With a three-year buffer, about 70% of the acknowledgment coefficient survives subtraction of the future-editor signal. We therefore center the interpretable connection evidence on acknowledgments, not editor tenure.

### *Threat 4: selection into the preprint frame*

Conditioning on NBER entry could mechanically inflate the connection coefficient. A Heckman selection correction using pre-2010 NBER working-paper history as the substantive exclusion yields a negative, significant inverse-Mills coefficient and raises the connection coefficient by 8.3%. Selection bias, if anything, runs downward here: the ordinary-least-squares estimate behaves like a floor, not a ceiling, reversing the usual prior about selection in this setting. (The Mills magnitude is sensitive to the exclusion; the functional-form-only estimate is larger and less credible than the exclusion-based one, which we report as the headline.)

***Threat 5: selection-on-observables fragility***

Propensity-score matching, coarsened exact matching, entropy balancing, and augmented inverse-probability weighting attenuate the connection estimate by 4.5% on average; tightening author identity resolution in OpenAlex changes the acknowledged-editor coefficient by 1.5%. The estimate is not an artifact of functional form. SI Figure S1 collects these as a sensitivity stack, with target variables, baselines, adjusted values, and percent changes reported separately because the checks are not the same estimator.

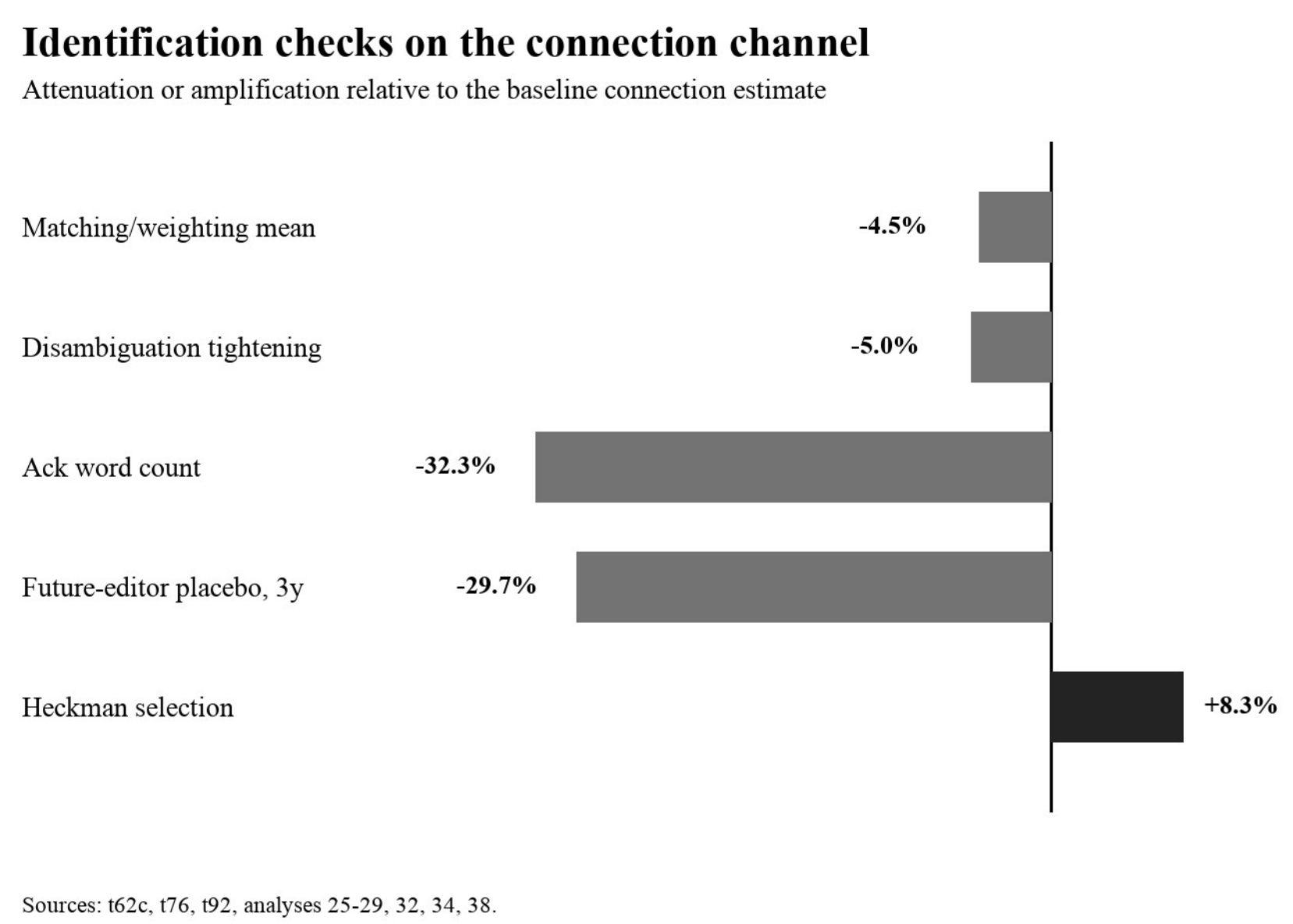

**Figure S1 | Identification checks on the connection channel**

***A design-based check, and its honest limit***

A natural question is whether editor rotations, the design used in causal studies of editorial networks (4), deliver a cleaner estimate. We run a tier-level difference-in-differences on ever-tied paper–tier cells. The active-tie coefficient is +0.046 (tier and year fixed effects) and survives idea and execution controls at +0.028, indicating a connection–placement association that is not just unmeasured quality. But the event study cannot certify a discrete jump at the moment an editor joins a board: the linear event-time trend is significant (+0.017) while the jump at entry is not (−0.011, n.s.). We therefore present the rotation result as corroborating a stock-of-connection effect that survives quality controls, not as a clean flow-at-entry causal design, and we do not lean on it for identification. The rotation design as specified is underpowered to separate a jump from a trend.

***Residual threat***

What we cannot close is the absence of randomized editor assignment and the possibility of unobserved paper-level quality orthogonal to all five blocks. The contribution is to bound, not eliminate, this threat, and to show that every credible adjustment leaves a positive, bounded, apex-concentrated connection gradient.

## S5. Robustness of the broad picture

The factor-share ordering and the dual channel are stable across the alternative outcomes (SI Table B1), across all 216 of 216 specifications in a specification curve (all same-signed, $p < 0.05$), and across cohorts.

Table B1. Shapley factor shares across all placement margins and outcomes. Visualized in Figure 2.

| DV | Block | Shapley R2 | Share | Block R2 |
|---|---|---|---|---|
| DV-A: Top-5 | SFT predicted idea quality | 0.0226 | 13.1% | 0.1732 |
| DV-A: Top-5 | LLM text score | 0.0240 | 13.9% | 0.1732 |
| DV-A: Top-5 | Execution | 0.0270 | 15.6% | 0.1732 |
| DV-A: Top-5 | Author ability | 0.0323 | 18.7% | 0.1732 |
| DV-A: Top-5 | Connections | 0.0672 | 38.8% | 0.1732 |
| DV-B: Prestige | SFT predicted idea quality | 0.0321 | 16.4% | 0.1959 |
| DV-B: Prestige | LLM text score | 0.0161 | 8.2% | 0.1959 |
| DV-B: Prestige | Execution | 0.0730 | 37.3% | 0.1959 |
| DV-B: Prestige | Author ability | 0.0272 | 13.9% | 0.1959 |
| DV-B: Prestige | Connections | 0.0475 | 24.2% | 0.1959 |
| DV-C: Top-field | SFT predicted idea quality | 0.0072 | 12.3% | 0.0581 |
| DV-C: Top-field | LLM text score | 0.0006 | 1.1% | 0.0581 |
| DV-C: Top-field | Execution | 0.0306 | 52.6% | 0.0581 |
| DV-C: Top-field | Author ability | 0.0086 | 14.8% | 0.0581 |
| DV-C: Top-field | Connections | 0.0112 | 19.2% | 0.0581 |
| DV-D: Mid+ | SFT predicted idea quality | 0.0205 | 15.7% | 0.1308 |
| DV-D: Mid+ | LLM text score | 0.0051 | 3.9% | 0.1308 |
| DV-D: Mid+ | Execution | 0.0654 | 50.0% | 0.1308 |
| DV-D: Mid+ | Author ability | 0.0153 | 11.7% | 0.1308 |
| DV-D: Mid+ | Connections | 0.0245 | 18.7% | 0.1308 |
| DV-E: Any publication | SFT predicted idea quality | 0.0053 | 9.5% | 0.0563 |
| DV-E: Any publication | LLM text score | 0.0013 | 2.2% | 0.0563 |
| DV-E: Any publication | Execution | 0.0347 | 61.5% | 0.0563 |
| DV-E: Any publication | Author ability | 0.0065 | 11.5% | 0.0563 |
| DV-E: Any publication | Connections | 0.0086 | 15.3% | 0.0563 |
| DV-F: Hurdle | SFT predicted idea quality | 0.1489 | 15.6% | 0.9566 |
| DV-F: Hurdle | LLM text score | 0.0660 | 6.9% | 0.9566 |
| DV-F: Hurdle | Execution | 0.4050 | 42.3% | 0.9566 |
| DV-F: Hurdle | Author ability | 0.1168 | 12.2% | 0.9566 |
| DV-F: Hurdle | Connections | 0.2200 | 23.0% | 0.9566 |
| DV-G: Citation-wtd prestige | SFT predicted idea quality | 0.0272 | 15.5% | 0.1750 |
| DV-G: Citation-wtd prestige | LLM text score | 0.0106 | 6.1% | 0.1750 |
| DV-G: Citation-wtd prestige | Execution | 0.0729 | 41.7% | 0.1750 |
| DV-G: Citation-wtd prestige | Author ability | 0.0185 | 10.6% | 0.1750 |
| DV-G: Citation-wtd prestige | Connections | 0.0458 | 26.2% | 0.1750 |
| DV-H: log citations | SFT predicted idea quality | 0.0095 | 10.1% | 0.0940 |
| DV-H: log citations | LLM text score | 0.0013 | 1.4% | 0.0940 |
| DV-H: log citations | Execution | 0.0575 | 61.1% | 0.0940 |
| DV-H: log citations | Author ability | 0.0081 | 8.7% | 0.0940 |
| DV-H: log citations | Connections | 0.0176 | 18.7% | 0.0940 |
| DV-lin: Placement | SFT predicted idea quality | 0.0279 | 15.3% | 0.1825 |
| DV-lin: Placement | LLM text score | 0.0130 | 7.1% | 0.1825 |
| DV-lin: Placement | Execution | 0.0757 | 41.5% | 0.1825 |
| DV-lin: Placement | Author ability | 0.0239 | 13.1% | 0.1825 |
| DV-lin: Placement | Connections | 0.0420 | 23.0% | 0.1825 |

## S6. Heterogeneity: fields, teams, and time

Three cuts show that the dual channel is general but not uniform.

*Across fields.* The connection gradient varies roughly twofold across the eight largest subfields: the no-to-high-connection rise in the apex rate is largest in public economics (+0.236), international (+0.235), labor (+0.218), and macroeconomics (+0.204), and smallest in theory (+0.163), econometrics (+0.145), finance (+0.144), and health (+0.141). We read this not as "some fields gatekeep more" but as a measurement artifact consistent with the acknowledgment-based placebo: the connection signal runs largely through written acknowledgments, and fields with denser acknowledgment cultures (macroeconomics, public, international) mechanically show larger measured gradients than fields with sparse ones (theory, finance). Execution heterogeneity points the same way: the execution dimension most correlated with prestige is writing and presentation in the theory-leaning fields and identification strategy in the applied empirical fields, so "what counts as good execution" is field-specific even though execution is the largest input everywhere.

*Across team size.* Favoritism compounds with coauthorship. Among fair-rated papers, the no-to-high-connection elevator is 2.1-fold for solo authors (10.4% to 21.9% prestige placement) but 3.7-fold for teams of three or more (16.3% to 61.0%), consistent with a mechanical channel: more coauthors mean more chances that at least one carries an editor tie, so a team draws on the union of its members' networks. (The high-connection, three-plus-author fair cell contains 59 papers, so we report this as descriptive, not estimated.) This is the team-production analog of the connection channel in the spirit of the increasing dominance of teams in knowledge production (29): teams do not just produce more knowledge, they aggregate more network capital.

*Across time.* The result is not an artifact of the early cohorts. Over 2010–2015 the apex quality bar rose modestly (mean execution of apex papers trended up by about 0.037 standard deviations per year) while the connection elevator stayed flat (the no-to-high gradient moved by −0.002 per year, indistinguishable from zero). The bar is tightening; the elevator is not eroding. A connection result driven by editorial idiosyncrasy in the earliest cohorts would show a declining gradient; it does not.

## S7. Data availability

A replication package containing de-identified data and the code that reproduces every table and figure is available at https://github.com/ln9527/publishing-production-function. Each row is keyed by the public NBER working-paper identifier and carries only numeric derived measures, scores, and outcomes; author and editor identities (names, identifiers, paper titles, editor names, and journal identities) are withheld for privacy. Because the measures are derived from public sources (the National Bureau of Economic Research, OpenAlex, Crossref, Semantic Scholar, and the published editorial boards of economics journals), the identifying layer can be rebuilt independently from those sources. No proprietary or terms-restricted source was scraped or mined. The repository is private pending publication and will be made public on acceptance.